\documentstyle[12pt]{article}
\begin{document}

{\Large
 The Nuclear Effect Induced by Additional Parton Evolution and Recombination in
    Nuclear Environment} 

\vspace{5mm}
\begin{center}
{\large Jian-Jun Yang$^{1,3}$, Guang-Lie Li$^{2,3}$} 
\end{center}

\vspace{5mm}
{\small 
$^1$Department of Physics, Nanjing Normal University, Nanjing 210024, China

$^2$CCAST(World Laboratory), P.O.Box 8730, Beijing 100080, China

$^3$Institute of High Energy Physics, Academia Sinica, P.O.Box 918(4), 
Beijing 100039, China\footnote{Mailing address. 
Email address: yangjj@bepc3.ihep.ac.cn}
}

\vspace{5mm}

\begin{abstract}

 The relationship between the parton
transverse momentum and longitudinal momentum, which is  obtained by 
analysing the hadron tensor in the deep inelastic scattering (DIS) process, 
indicates that the transverse size of partons in a nucleon is sensitively
affected by the effective nucleon mass. The change in this transverse
size gives rise to an Additional Parton Evolution (APE) in the nuclear 
environment refer to 
the normal QCD parton evolution thus resulting in new parton distributions.
The self-consistent integral equations for these new parton distributions
are given. Having undergone the normal QCD evolution and
the additional evolution in the nuclear environment, the partons with the small 
$x$ participate in recombination and contribute to the nuclear shadowing effect.
The reasonably good descriptions of  the nuclear effect of 
structure functions, nuclear Drell-Yan ratios and nuclear gluon distributions
are  presented. A  quantitative comparison between the APE model and 
the $Q^2$-rescaling model is made.

\end {abstract}


\newpage

\pagestyle{plain}

\section{Introduction}
 Since the discovery of the nuclear modification in the nucleon 
structure function 
$F_2(x,Q^2)$ by the European Muon Collaboration (EMC) \cite {EMC-1},  
various models
have been proposed to investigate the nuclear effect. Several years 
ago, the New Muon
Collaboration (NMC) \cite {NMC-EMC} accurately measured the structure functions
$F_2$  at very small $x$. These data as well as the EMC 
data \cite {EMC-2} provide an insight for investigating the physics details
in the shadowing region($x<0.1$). The need for testing  
the  models  of  the EMC effect becomes more and more important for further 
understanding of the nature of the nuclear effect. Bickerstaff et al.
 \cite {Bickerstaff} found that although most of the theoretical
 models can provide 
 good explanations for the EMC effect, they can not describe 
the nuclear Drell-Yan process well. The NMC \cite {NMC} analyzed the inelastic
$J/\psi$ production data and obtained the gluon distribution ratios 
$G_{Sn}(x)/G_{C}(x)$ which indicate a modification of the gluon 
distribution in the nuclear environment. Kumano \cite {Kumano} has made progress 
 towards
the  unified description of the EMC effect and the nuclear gluon distribution in
the  $Q^2$-rescaling  model with the recombination effect. Although the $Q^2$-rescaling
model is a useful and effective model in explaining deep-inelastic data 
in the valence dominated medium $x$ region, it is not clear whether the 
momentum distributions of sea 
quarks and gluons follow the same rescaling as that  for valence 
quarks in the nuclear environment. According to the QCD theory, sea quarks have 
different dynamic properties while interacting with gluons. There is 
a constraint  condition on the  number conservation for 
valence quarks but not for sea quarks. So it is reasonable to believe
that the modification of sea quark distributions should be different   
from that of valence quarks. For further discussion, 
 we assume that  the transverse size of partons in a nucleon 
can be described by an average parton transverse momentum which is related 
to the longitudinal distributions of partons. 
 The relationship between the parton
transverse momentum and longitudinal momentum, which is  obtained by 
analysing the hadron tensor in the deep inelastic scattering process, 
indicates that the transverse size of partons in a nucleon depends sensitively
on the  effective nucleon mass. The change of this transverse
size gives rise to an additional parton evolution  in the nuclear environment refer to 
the normal QCD parton evolution thus resulting in new parton distributions. The probabilities
of the additional evolution  are directly related to the ratio of
the average parton transverse momentum in the bound  nucleon to
that in the free nucleon.
According to these probabilities, we give the self-consistent integral 
equations for parton distribution modifications 
due to the additional evolution in the nuclear environment. 
  
Considering the fact that the nucleon diameter is about 1.8$fm$, in an  
infinite momentum frame, one can estimate that the longitudinal size of a nucleon 
in a Lorentz contracted nucleus  is

 $$  D=1.8(fm)m_N/p_N,$$
and the longitudinal localization size of a parton with the momentum $xp_N$ is
$L=1/(xp_N)$. If a parton has a small $x$ and consequently its dimension ($L$) 
exceeds the nucleon longitudinal size ($D$),
it leaks out of the nucleon and recombines with  partons of other 
nucleons. We suppose that  the partons  participating in recombination
have undergone the normal QCD evolution and the 
additional evolution in the nuclear environment before the fusion. After considering 
the additional evolution in the nuclear environment  with the parton 
recombination effects, we find that the new model can give a well explaination 
to the  nuclear  
effects of the structure function, nuclear Drell-Yan process
 and inelastic $J/\psi$ production.  
Finally, a  quantitative comparison between the APE model and 
the $Q^2$-rescaling model is made.

\section{ Additional Parton Evolution(APE) Due to Nuclear Environment}

Partons in a free nucleon have their own distributions with a specified scale.
It is generally  accepted that the transverse size of partons in 
a nucleon is affected by the nuclear environment. We introduce an average 
transverse momentum to describe the change 
of the transverse size of partons.

In order to obtain the average transverse momentum of partons, let us repeat the 
$\xi$ variable analysis \cite {Gluck} of 
the hadron tensor $W_{\mu\nu}$ in the deep inelastic scattering 
process shown in Fig.1 . $W_{\mu\nu}$ can be written as
\begin{eqnarray}
W_{\mu\nu}&=&\frac{1}{2\pi}\int d^4x e^{iq{\cdot}x}\langle p|[{\bf J}_{\mu}(x),
{\bf J}_{\nu}(0)]|p \rangle \nonumber \\
	  &=&(-g_{\mu\nu}+\frac{q_{\mu}q_{\nu}}{q^2})W_1
	   + \frac{1}{m_N^2}(p_{\mu}-\frac{p\cdot q}{q^2}q_{\mu})
	(p_{\nu}- \frac{p \cdot q}{q^2}q_{\nu})W_2
\end{eqnarray}
where $m_N$ is the mass of a nucleon. If we take the frame in which 
$p=(m_N,{\bf 0})$ and neglect the quark mass, then 
\begin{equation}
 W_{\mu\nu} \propto \int \frac{d^3k}{2k_0}f(\frac{2k_0}{m_N})\delta(q^2+2k\cdot q)
w_{\mu\nu} \label {tensor}
\end{equation}
where 
\begin{equation}
w_{\mu\nu}=[(k+q)_\mu k_\nu + (k+q)_\nu k_\mu+\frac{q^2}{2}g_{\mu\nu}]
\end{equation}
is the quark tensor in the lab. frame and $f(k\cdot p)$ is a 
Lorentz-invariant wave  function. 
To extract $W_1$ and $W_2$ from $W_{\mu\nu}$, we introduce two contractions
\begin{equation}
 B=g^{\mu\nu}W_{\mu\nu}
=-3W_1+(1+\frac{{\nu}^2}{Q^2})W_2 \label {b}
\end{equation}
\begin{equation} 
 C=\frac{1}{{m_N}^2(1+{\nu}^2/{Q^2})}p^{\mu}p^{\nu}W_{\mu\nu}
=-W_1+(1+\frac{{\nu}^2}{Q^2})W_2 \label {c}
\end{equation}
and then $W_1$ and $W_2$ can be expressed as 
\begin{equation}
W_1=\frac{1}{2}(C-B)
\end{equation}
\begin{equation}
W_2=\frac{3C-B}{2(1+{\nu}^2/Q^2)}    
\end{equation}
If we take the frame in which
$q=(\nu,0,0,\sqrt{{\nu}^2+Q^2})$ and $k=(k_0,0,k_0\sin\theta,k_0\cos\theta)$,  
then
\begin{equation}
{k_\perp}^2=k_0^2 \sin^2\theta=\frac{2k_0^2+2k_0\nu-\frac12 Q^2}
{2(1+{\nu}^2/Q^2)}
\end{equation}
By means of Eq.(\ref {tensor}) and the contractions defined by
Eq.(\ref {b}) and Eq.(\ref {c}), one can find the relation
\begin{equation} 
\langle k^2_\perp \rangle = -\frac{CQ^2}{2B}
\end{equation}
Taking $W_1$ and $W_2$ in Eq.(\ref {b}) and 
Eq.(\ref {c}) as those in Refs.\cite {Georgi} and \cite {Barbieri} and 
neglecting $O(m_N^2/Q^2)$ and  
$O(Q^2/{\nu}^2)$ terms, one easily  obtains 
\begin{equation}
{\langle K_{\perp}^2(x,Q^2) \rangle}^{{q}^{N(A)}}=\frac{x^3m_{N(A)}^2}
{xq^{N(A)}(x,Q^2)}\int_x^1 dx^{\prime } 
\frac{x^{\prime }q^{N(A)}(x^{\prime },Q^2)}{x^{\prime ^2}} \label {kq}
\end{equation}
Furthermore, we assume that there is a similar relationship between
the gluon transverse momentum and  longitudinal momentum distributions, i.e.,
\begin{equation}
{\langle K_{\perp}^2(x,Q^2)\rangle}^{{G}^{N(A)}}=\frac{x^3m_{N(A)}^2}
{xG^{N(A)}(x,Q^2)}\int_x^1dx^{\prime } 
\frac{x^{\prime }G^{N(A)}(x^{\prime },Q^2)}{x^{\prime ^2}} \label {kg}
\end{equation}
Eq.(\ref {kq}) and Eq.(\ref {kg}) with the bound nucleon effective  mass 
$m_{N(A)}$ $(m_{N(1)}=m_N)$ have been written in a unified form 
for partons in both a free nucleon(N) and a bound nucleon(A) in a 
nucleus with the mass number $A$.
The effective mass of the bound nucleon  
is less than that of the free one, i.e., in terms of Eq.(\ref {kq}) and 
Eq.(\ref {kg}), the average transverse momentum of partons in the 
bound nucleon is less than that in the free nucleon, 
which mainly causes the
parton evolution in the nuclear environment. By using the same technique in deriving 
the famous Altarelli-Parisi evolution \cite {A-P}, we obtain    
the evolution probabilities for the evolution from a
parton with momentum fraction $y$ to another parton with the 
momentum fraction $x$; 
\begin{equation}
F_{q_i\rightarrow q_j}(\frac xy)=\frac{\alpha_s}{2\pi}C_2(R)
[\frac{1+(\frac xy)^2}
{(1-\frac xy)_{+ }}+\frac 32\delta (1-\frac xy)]
ln \frac{{\langle K_{\perp}^2(x,Q^2) \rangle}^{q_j^N}}
{{\langle K_{\perp}^2(x,Q^2) \rangle}^{q_j^A}}
\end{equation}
\begin{equation}
F_{G\rightarrow q_j}(\frac xy)=\frac{\alpha _s}{4\pi}[(\frac xy)^2+
(1-\frac xy)^2]ln \frac{{\langle K_{\perp}^2(x,Q^2)\rangle}^{q_j^N}}
{{\langle K_{\perp}^2(x,Q^2)\rangle}^{q_j^A}}
\end{equation}
\begin{equation}
F_{q_i\rightarrow G}(\frac xy)=\frac{\alpha _s}{2\pi}C_2(R)
[\frac {1+(1- \frac xy)^2}{\frac xy}]
ln \frac{{\langle K_{\perp}^2(x,Q^2)\rangle}^{G^N}}
{{\langle K_{\perp}^2(x,Q^2)\rangle}^{G^A}}
\end{equation}
$$F_{G\rightarrow G}(\frac xy)=\frac{\alpha _s}{\pi}C_2(G)
[\frac {\frac xy}{(1- \frac xy)_{+}}+\frac {(1- \frac xy)}{\frac xy}
+\frac xy(1-\frac xy)$$
\begin{equation}
+\frac {1}{12}(11-\frac{2N_f}{3})\delta(1-\frac xy)] 
ln \frac{{\langle K_{\perp}^2(x,Q^2)\rangle}^{G^N}}
{{\langle K_{\perp}^2(x,Q^2)\rangle}^{G^A}}
\end{equation}
where the strong interaction coupling constant $\alpha_s$ is 
$$\alpha_s(Q^2)=4\pi/[\beta_0ln(Q^2/\Lambda^2)]$$
with $\Lambda=0.19GeV$ and $\beta_0=11-2N_f/3$,$N_f=3$.
$C_2(R)=\frac 34$ and $C_2(G)=3$ are color factors for $N_c=3$. 
Therefore, the quark distributions modified by the additional parton 
evolution(APE) in the nuclear environment can be written as follows:
\begin{equation}
q_v^{APE}(x,Q^2)=q_v^N(x,Q^2)+\int_x^1F_{q_v\rightarrow
q_v}(\frac xy)\frac{q_v^N(y,Q^2)}ydy
\end{equation}
$$ q_s^{APE}(x,Q^2)=q_s^{N}(x,Q^2)+\int_x^1F_{q_s\rightarrow
q_s}(\frac xy)\frac{q_s^{N}(y,Q^2)}ydy $$
\begin{equation}
+\int_x^1F_{G\rightarrow q_s}(\frac xy)\frac{G^{N}(y,Q^2)}ydy
\end{equation}
$$ G^{APE}(x,Q^2)=G^{N}(x,Q^2)+\sum \limits_{i}\int_x^1F_{q_i\rightarrow
G}(\frac xy)\frac{q_i^{N}(y,Q^2)}ydy $$
\begin{equation}
+\int_x^1F_{G\rightarrow G}(\frac xy)\frac{G^{N}(y,Q^2)}ydy
\end{equation}                                               
where the input parton distributions $q^N(x,Q^2)$ and $G^N(x,Q^2)$ 
are taken from 
Refs. \cite {KMRS}- \cite {D-O}. 
In fact, these equations are the self-consistent  integral equations since 
the evolution probabilities $F_{p_i\rightarrow p_j}(z)$ contain the
parton transverse momenta which are related to the   
parton longitudinal momentum distributions $q^{N(A)}(x,Q^2)$ and 
$G^{N(A)}(x,Q^2)$ by 
Eq.(\ref {kq}) and Eq.(\ref {kg}).
In the calculation of the parton distributions in the bound nucleon, the 
numerical solution of these equations is obtained by the iteration 
method.

\section{ Leak-out  Sea Quarks and Gluons in Nuclear Environment}

Although the leakage in the nuclear environment can occur for all partons, the
most important contribution arises from partons with the largest 
spatial uncertainty, i.e., those with the small $x$.
 We assume that  the leak-out  partons are  
 sea quarks and  gluons for which  the momentum cutoff function 
 \cite {Close}
is taken as
  $$ \beta(x)=exp(-\frac {m_N^2y_0^2x^2}2),$$
      namely, the distributions of leak-out(LK) partons  are
     $$ p^{LK}(x,Q_0^2)=\beta(x)p^{APE}(x,Q_0^2)  $$
where  the input parton distributions  
$p^{APE}=u_{s}^{APE}$, ${\bar{u}}^{APE}$, $d^{APE}_s$, ${\bar{d}}^{APE}$, 
$s^{APE}$ , ${\bar{s}}^{APE}$ and $G^{APE}$ together with 
$u^{APE}_v$ and $d^{APE}_v$  are those having 
undergone the normal QCD parton evolution and the additional parton 
evolution due to the nuclear environment.
The momentum cutoff function is shown as a function of $x$ in Fig.2 with 
various $y_0$. As discussed in section 1 , the partons with $x \leq 0.11$
 can leak out of a nucleon. We then find in Fig.2 that an appropriate
choice of  $y_0$ is 2.0$fm$.

\section{ Parton Recombination and Further Modification to Parton
	    Distributions}

 According to the original idea of the parton 
recombination by Close, Qiu and  Roberts \cite {Close}, 
a leak-out parton can fuse with a parton from another nucleon.
 In general, the
modification of a parton distribution $p_3(x_3)$, due to the process of 
producing the parton $p_3$ with the momentum $x_3$ by fusion of partons
  $p_1$ and $p_2$, is given by \cite {Close},
$$\Delta p_3(x_3)=K \int dx_1dx_2p_1(x_1)p_2(x_2)$$
   $$ \times \Gamma _{p_1p_2\rightarrow p_3}(x_1,x_2,x_3=x_1+x_2)\delta(x_3,x_1,x_2),$$ 
where $K$ is given by Refs. \cite {Close} and \cite {Mueller},
   $$ K=9A^{1/3}\alpha_s/(2{R_0}^2Q_0^2) $$
The nuclear radius  is $R=R_0A^{1/3}$ with $R_0=1.1$ $fm$ \cite {Barrett}. 
 The  function $\delta(x_3,x_1,x_2)$ defined by Ref. \cite {Kumano}
is introduced  to keep the momentum conservation. The parton fusion 
function $\Gamma_{p_1p_2\rightarrow p_3}(x_1,x_2,x_3)$ is a probability 
for producing a parton $p_3$
with momentum $x_3$ by a fusion of partons $p_1$ and $p_2$ with momentum 
$x_1$ and $x_2$, respectively. It is  related to a splitting function
$P_{p_1 \leftarrow p_3}(z)$ in the Altarelli-Parisi equations \cite {A-P} by
\begin{equation}
 \Gamma_{p_1p_2\rightarrow p_3}(x_1,x_2,x_3)=\frac{x_1x_2}{x_3^2}
P_{p_1\leftarrow p_3}[\frac{x_1}{x_3}]C_{{p_1}{p_2}\rightarrow p_3}. 
\end{equation}
where $C_{p_1 p_2\rightarrow p_3} $ is the ratio of color factors in the 
processe 
$p_1p_2\rightarrow p_3$. Using the fusion functions, one can  obtain the 
modifications 
 of parton distributions due to the recombination(RC)\cite {Kumano}. 
With our notations, the explicit expressions for the modifications 
$\Delta q^{RC}(x)$ and $\Delta G^{RC}(x)$ are

$x\Delta q^{RC}(x)$ 

\begin{eqnarray}
&=&  \frac{K}{6} \int_0^x
 \frac{dx^\prime}{x^\prime}
x^{\prime}(x-x^\prime)[G^{LK}(x^\prime)q^{APE}(x-x^\prime) \nonumber \\
 & &+ G^{APE}(x^\prime)q^{LK}(x-x^\prime)][1+{[
\frac{x-x^\prime}{x}]}^2] \nonumber \\
 & &- \frac{K}{6} \int_0^1
\frac{dx^\prime}{x^\prime}
xx^{\prime}[q^{APE}(x)G^{LK}(x^\prime)+q^{LK}(x)G^{APE}(x^\prime)]
\frac{x}{x+x^\prime}[1+{[ \frac{x}{x+x^\prime}]}^2] \nonumber \\
 & &- \frac{4K}{9}x
\int_0^1dx^\prime x x^\prime
[q^{APE}(x){\bar{q}}^{LK}(x^\prime)+q^{LK}(x){\bar{q}}^{APE}(x^\prime)]
\frac{x^2+{x^\prime}^2}{{(x+x^\prime)}^4}
\end{eqnarray}
and

$x\Delta G^{RC}(x)$ 
\begin{eqnarray}
 &=&  \frac{3K}{4}
x \int_0^x dx^\prime x^\prime (x-x^\prime)
[G^{APE}(x^\prime)G^{LK}(x-x^\prime) \nonumber  \\
 & &+G^{LK}(x^\prime)G^{APE}(x-x^\prime)]
\frac{1}{x^2}[\frac{x^\prime}{x-x^\prime}+ \frac{x-x^\prime}{x^\prime}+
 \frac{x^\prime (x-x^\prime)}{x^2}] \nonumber \\
 & &- \frac{3K}{4}x
 \int_0^1 dx^\prime x x^\prime
[G^{LK}(x)G^{APE}(x^\prime)+G^{APE}(x)G^{LK}(x^\prime)] \nonumber \\
 & &\times \frac{1}{(x+x^\prime)^2}  
[ \frac{x}{x^\prime}+ \frac{x^\prime}{x}+
\frac{xx^\prime}{(x+x^\prime)^2}] \nonumber  \\
 & & + \frac{4K}{9}x\int_0^xd{x}^\prime  
 \sum \limits_{i}[x^\prime q_i^{APE}(x^\prime)(x-x^\prime) 
{\bar{q}}_i^{LK}(x-x^\prime) \nonumber \\
 & & +x^\prime \bar{q}_i^{APE}(x^\prime)(x-x^\prime) 
q_i^{LK}(x-x^\prime)]\frac{1}{x^4}[{x^\prime}^2+(x-x^\prime)^2] \nonumber \\
 & & - \frac{K}{6}
 \int_0^1dx^\prime x G^{APE}(x)
\sum \limits_{i}[x^\prime q_i^{LK}(x^\prime)
+x^\prime {\bar{q}}_i^{LK}(x^\prime)]
\frac{1}{x+x^\prime}[1+[
\frac{x^\prime}{x+x^\prime}]^2] \nonumber \\
 & & - \frac{K}{6}
 \int_0^1dx^\prime x G^{LK}(x)
\sum \limits_{i}[x^\prime q_i^{APE}(x^\prime)
+x^\prime {\bar{q}}_i^{APE}(x^\prime)]  \nonumber \\
 & &\times \frac{1}{x+x^\prime}[1
+{[\frac{x^\prime}{x+x^\prime}]}^2]
\end{eqnarray}
The explicit $Q^2$ dependence in the parton distributions is not shown in the above
two equations in order to simplify the notation. 
Therefore, we write, for the nuclear gluon distribution,
\begin{equation}
xG^{A}(x,Q_0^2)=xG^{APE}(x,Q_0^2)+\Delta xG^{RC}(x,Q_0^2)
\end{equation}
Similarly, for the nuclear quark distribution,
\begin{equation}
xq^{A}(x,Q_0^2)=xq^{APE}(x,Q_0^2)+ \Delta xq^{RC}(x,Q_0^2)+ 
\Delta xq^{GS}_{s}(x,Q_0^2)
\end{equation}
where $\Delta xq^{GS}_{s}(x,Q_0^2)$ is the modification due to 
the gluon shadowing(GS) and is derived from the evolution equations. A crude
estimate of $\Delta xq_{s}^{GS}(x,Q_0^2)$ is given by \cite {Close}
\begin{equation}
xq_{s}(x,Q_0^2)=-\frac{x}{12}\frac{\partial [xG(x,Q_0^2)]}{\partial x} \label {SDQ}
\end{equation}
Using this relation, we expect that the sea-quark distribution 
is also affected by the gluon modification as follows:
\begin{equation}
x\Delta q^{GS}_{s}(x,Q_0^2)=-\theta(x_0-x)\frac{x}{12}
\frac{\partial [x\Delta G^{RC}(x,Q_0^2)]}{\partial x}   
\end{equation}
where the step function $\theta (x_0-x)$ is introduced because the relation 
is valid only at very small $x$ ($x$ less than $x_0$ with $x_0$=0.1).

\section{Explanation of Nuclear Effect in  Nucleon Parton Distributions}

 We have considered the additional parton 
evolution and parton recombination in sections 2 and 4, 
respectively, and 
gotten the nuclear parton distributions at $Q_0^2=4.0GeV^2$. To compare 
the results with the experimental data or display the $Q^2$-dependence of the 
model, these
distributions are evolved to those at lager $Q^2$ by
using the ordinary Altarelli-Parisi equations \cite {A-P}.
For the valence quark($q_{v}$), sea quark($q_{s}$), and gluon(G), the evolution 
equations are
\begin{equation}
\frac{\partial}{\partial t}q^{A}_v(x,t)=\int_x^1 \frac{dy}{y}
q_v^{A}(y,t)P_{qq}[\frac{x}{y}]
\end{equation}
\begin{equation}
\frac{\partial}{\partial t}q^{A}_s(x,t)=\int_x^1 \frac{dy}{y}
[q_s^{A}(y,t)P_{qq}[\frac{x}{y}]+G^{A}(y,t)P_{qG}[\frac{x}{y}]]
\end{equation}
\begin{equation}
\frac{\partial}{\partial t}G^{A}(x,t)=\int_x^1 \frac{dy}{y}
[q_s^{A}(y,t)P_{Gq}[\frac{x}{y}]+G^{A}(y,t)P_{GG}[\frac{x}{y}]]
\end{equation}
where $P_{ij}$ is the splitting function, and $t$ is defined by
\begin{equation}
t=-(2/\beta_0)\ln [\alpha_s(Q^2)/\alpha_s(Q_0^2)]
\end{equation}
We used the solution of these integral  equations in the leading order 
and the evolution subroutine of Ref. \cite {K-L} in our numerical calculation. 

\subsection{Nuclear Structure Function $F_2^A(x)$}

We now proceed to compute the modification in the  structure 
function.
In order to compare  with the 
experimental  data , 
let us  define the ratio of the average nuclear structure function to the
deuteron structure function as \cite {Li1}
\begin{equation} 
R^{A/D}(x,Q^2)=\frac{F_2^A(x,Q^2)}{F_2^D(x,Q^2)}\label {RAD}
\end{equation}
with the average nuclear structure function:
\begin{equation} 
F_2^A(x,Q^2)=\frac 1A[F_{2A}(x,Q^2)-\frac 12(N-Z)(F_2^n(x,Q^2)-F_2^p(x,Q^2))]
\end{equation}
where $F_2^n$ and $F_2^p$ are the free neutron and proton structure
functions, respectively. $F_{2A}(x,Q^2)$ is the nuclear structure function. 
The second term compensates for the neutron excess.
By considering the Fermi motion of nucleons in the nucleus, 
$F_{2A}(x,Q^2)$ can be written as \cite {Li1}
\begin{equation}
F_{2A}(x,Q^2)=\sum \limits_\lambda\int \frac {d^3p}{(2\pi)^3}
\left|\psi_{\lambda}(\vec{p})\right|^2 zF_2^{N(A)}(\frac {x}{z},Q^2),
\end{equation}
with $z=(p_0+p_3)/m_N,p_0=m_N+{\epsilon}_\lambda$ , $\epsilon_\lambda$
 is the separation energy of a nucleon in the single-particle
 state $\lambda$, and $\psi_{\lambda}(\vec{p})$ is the 
single-particle wave function of the nucleon in the momentum space, which 
satisfies the light-cone normalization:
\begin{equation} 
 \int\frac{d^3p}{(2\pi)^3}\left|\psi_{\lambda}(\vec{p})\right|^2z=1 
\end{equation}
In the following calculation,  $\epsilon_\lambda$ and  
$\psi_{\lambda}(\vec{p})$ 
are taken from Ref. \cite {Li2}. By using the nuclear parton distributions
after considering the additional parton evolution, recombination and  
the ordinary Altarelli-Parisi evolution to higher $Q^2$, the bound nucleon structure function
$F_2^{N(A)}(x,Q^2)$ can be expressed as
$$ F_2^{N(A)}(x,Q^2)$$
\begin{eqnarray}
&=& \frac 1{18}x\{5[u_v^A(x,Q^2)+
d_v^A(x,Q^2)+u_s^A(x,Q^2)+
\bar{u}^A(x,Q^2) + 
d_s^A(x,Q^2) \nonumber \\
& & +\bar{d}^A(x,Q^2)]+2[s^A(x,Q^2)+
\bar{s}^A(x,Q^2)]\}
\end{eqnarray}
We now calculate $R^{A/D}(x,Q^2)$ in Eq.(\ref {RAD}).  The deuteron structure 
function $F_2^D(x,Q^2)$ is taken from Ref.\cite {Frankfurt}. Similar to 
Ref.\cite {Hendry}, we take 
\begin{equation}
m_{N(A)}=\frac{1}{1+0.48\ln(2-A^{-1/3}]}
\end{equation}
where the constant 0.48 is determined by $m_{N(\infty)}=0.75m_N$ \cite {GEB}.

The theoretical result in the range $0.1<x<0.2$ can not be well
described since the gluonic modification due to Eq.( \ref {SDQ}) is valid
at very small $x$. However, using the KMRS-B0\cite {KMRS}, 
our calculation results indicate that the modified 
parton distribution can be well parametrized by the analytical form of
the corresponding input distribution. The parameters in the analytical
form of the modified parton distributions can be obtained by fitting
the numerical results. So we obtain a smooth curve of
the ratio $R^{Ca/D}(x,Q^2)$ which is
 compared with
 the experimental data of the EMC effect
\cite {EMC-1} in Fig.3. The 
result  is in qualitative 
agreement with the experimental data. To show the distinct effect of
the additional parton evolution, the following results are no longer  
presented by fitting numerical results and taking into account the
Fermi-motion correction.

In order to investigate  the $Q^2$-dependence of the ratio $R^{Ca/D}(x,Q^2)$, the 
results obtained by using the GRV inputs\cite {GRV}
and the KMRS-B0 inputs\cite {KMRS} are plotted in Fig.4(a) 
and Fig.4(b), respectively.
One can easily find that the result 
obtained by using the KMRS-B0 inputs is less dependent on $Q^2$.
In Fig.5, the results with various 
input parton distributions of different parameterizations are presented, and they    
 show that, in the small $x$ region, the recombination results after
 undergoing the additional parton evolution are very sensitive to the input
sea-quark and gluon distributions. Using the Duke-Owens(2) 
inputs of Ref. \cite {D-O} 
with hard gluon distributions, theoretical
results underestimate the shadowing, which is  
not difficult to be understood since the parton parameterizations of
 Ref. \cite {D-O} did not fit to the experimental data in the small $x$ 
region. This is the reason why we take KMRS-B0 \cite {KMRS} or 
GRV \cite {GRV} parameterizations as input parton distributions in  
most of the following calculations.

\subsection{Nuclear Drell-Yan Ratio}

Several years ago, the E772 Collaboration \cite {Alde}
at Fermilab published the data of high-mass
dilepton production measured in the nuclear Drell-Yan (DY) process.
 These data aroused special attention in clarifying 
the different explanations for the nuclear effect  
on  the parton distributions.
Their results show that the ratio of DY dimuon yield per 
nucleon on a nuclear target to that on a free nucleon is slightly 
less than unity if the momentum fraction $x$ carried by 
a target quark is less than 0.1. The ratios over the range 
$0.1 < x < 0.3$, however, do not reveal distinct nuclear dependence.

In order to show the prediction of nuclear Drell-Yan ratio in our 
present model, let us begin
with the definition of the nuclear Drell-Yan ratio :
\begin{equation} 
T^{A/D}(x)=\frac{\int dx^{\prime} d^{2}\sigma^{h-A}(x^{\prime},x)/dx^{\prime}dx}
{\int dx^{\prime} d^{2}\sigma^{h-D}(x^{\prime},x)/dx^{\prime}dx},\label {DYR}
\end{equation}
where  the differential cross section of the nuclear Drell-Yan process  is:
\begin{equation}
\frac{d^{2}\sigma^{h-A}}{dx^{\prime}dx}={1 \over 3}\frac{4\pi\alpha^{2}}
   {3x^{\prime}xM^{2}}H_{h-A}(x^{\prime},x)
\end{equation}
with 
\begin{equation}
H_{h-A}(x^{\prime},x)=\sum \limits_{i} e_{i}^{2}[x^{\prime}q_{i}^{h}(x^{\prime})x
\bar{q}_{i}^{A}(x)+
x^{\prime}\bar{q}_{i}^{h}(x^{\prime})xq_{i}^{A}(x)]
\end{equation}
In a more specified case for the proton-nucleus reaction:
$$ H_{p-A}(x^{\prime},x,Q^2)
 =\frac1{54}x^{\prime}[4u_v^N(x^{\prime},Q^2)+d_v^N(x^{\prime},Q^2)]xS^A(x,Q^2) $$
$$ +\frac{1}{54A}x^{\prime}S^N(x^{\prime},Q^2)x[(A+3Z)u_v^A(x,Q^2) $$
\begin{equation}
+(4A-3Z)d_v^A(x,Q^2)+2AS^A(x,Q^2)] 
\end{equation}
where the total sea quark is  $S=u_s+\bar{u}_s+d_s+\bar{d}_s+s+\bar{s}$. 
By using  the nuclear parton distributions 
in our model, we calculated the nuclear Drell-Yan ratios for 
$^{40}Ca$. 
The  integral range for $x^{\prime}$ in Eq.(\ref {DYR}) is determined according to
 the kinematic region  of  the
 experiment in  Ref.\cite {Alde}, i.e. 
 $x^\prime-x>0$, and $0.025 \leq x \leq 0.30$.
To avoid the uncertainty derived by 
the crude estimate in Eq.( \ref {SDQ}) for taking into 
account the gluonic modification, 
the results, except those in the  range $0.1 < x < 0.2$, are 
compared with the E772 \cite {Alde} data in Fig.6. 
Fig.6 indicates that the APE model as well as the $Q^2$-rescaling model
can explain the Drell-Yan ratio reasonably well. However, in the small $x$
region($x < 0.1$), the result obtained by the APE model seems to be better  
than that of the $Q^2$-rescaling model.  

\subsection{ Nuclear Gluon Distribution}

In 1992, the NMC \cite {NMC} analyzed the inelastic  $J/\psi$ production result
with $R_{in}(Sn/C)=1.13\pm0.08$ by the color singlet(CS) model \cite {Berger} and
obtained the  gluon distribution ratios  $G_{Sn}(x)/G_{C}(x)$. The 
result sheds light on gluon distributions in nuclei.

The results of $G_{Sn}(x)/G_{C}(x)$ calculated in our model
together with the experimental data from NMC are presented in Fig.7, which
are qualitatively similar to those given by the $Q^2$-rescaling model 
in Ref.\cite {Kumano}. Due to the gluon shadowing, 
the ratios  $G_{Sn}(x)/G_{C}(x)$ are less than  1  at 
$x< 0.09$. It is noteworthy  that experimental errors are very large in 
comparison with typical theoretical modifications. In order to test 
our model, we should wait for more accurate  measurements of $G_A(x)$, 
for example, a proposed experiment at RHIC \cite{RHIC} in 
the small $x$ region for investigating details of the gluon shadowing.

 \section{Discussion and Summary}

It is worthwhile to compare the results of the present model with those of
the $Q^2$-rescaling model. Although these two models seem to have the same 
physics origin, they have  different  descriptions  of the nuclear  
effect on sea quarks and gluons. In the APE model, the evolution of valence 
quarks, sea quarks and gluons are 
determined by different interaction vertex 
factors, and their effects on their own momentum distributions are 
also different. However, the $Q^2$-rescaling model, ignoring the  
 difference in additional parton evolution modes for 
 different kinds of partons, 
 describes the influence of the nuclear environment on the 
 distribution functions of valence and sea quarks by using 
 the same $Q^2$-rescaling mechanism. In the medium $x$ region where the 
 valence quarks dominate, the $Q^2$-rescaling model can explain
 the EMC effect reasonably well, indicating that the nuclear 
 effect on the valence quark momentum distribution can indeed be well  
 described by the $Q^2$-rescaling mechanism. In the small $x$ region
 where sea quarks and gluons dominate, it is not clear 
 whether the $Q^2$-rescaling picture could be used  
 also for sea quarks and gluons. To avoid this uncertainty, we attempt to
 find a unified description of the nuclear effect on sea quarks and gluons 
as well as valence quarks. This is the purpose of our investigation.
 To make a quantitative comparison between the APE model and 
the $Q^2$-rescaling model, 
some results of these two models are shown in Figs.8-9 in 
different $x$ scales. 
Firstly, Figs.8-9 indicate that both models give reasonable 
explanations of the EMC effect in the 
medium $x$ region. However, the expectations of the 
ratio $R^{Ca/D}$ for two different 
input distributions (GRV and Duke-Owens) are almost the same in the APE model 
but distinctly different in the $Q^2$-rescaling model in which the ratio is  
slightly sensitive to the choice of free parton distributions.
In the lager $x$ region, there are distinct modifications of the recombination
effect in the ratio of $R^{Ca/D}$ which are similar to,
but different from  the contributions of 
the nuclear Fermi motion (cf. Fig.3).
For the small $x$ region, we are always seeking for some mechanisms 
which account for the
nuclear shadowing effect. Unfortunately, neither the APE model 
nor the $Q^2$-rescaling model can give suitable explanation of the 
nuclear shadowing effect. It has been 
emphasized by Ref.\cite {Kumano} that the shadowing due to the recombinations 
in the nuclear environment is mainly produced as a result of the modification 
in the gluon distribution. Considering this recombination  effect, both the 
$Q^2$-rescaling model (see Ref. \cite {Kumano}) and the APE model (see Fig.4) 
can fairly explain the EMC and NMC shadowing data  by using the KMRS-B0 input 
distributions. But, by using harder input gluon distributions such as 
those of the  Duke-Owens(2)\cite {D-O}, 
the theoretical results (see Figs.8-9)
 of these two models underestimate the shadowing.
 It seems, however, that by considering  the recombination effect, the 
prediction of nuclear shadowing in the APE model 
is closer to the experimental data than that in the $Q^2$-rescaling model with
these harder input gluon distributions. This is possibly  attributed to
the better descriptions of the nuclear 
effect on sea quarks and gluons in our model.

 As for the prediction of the nuclear Drell-Yan ratio, the result in Fig.6 
indicates 
that, with the same input parton distributions of GRV, the APE model has  
almost the same prediction of the nuclear Drell-Yan ratio
for $^{40}Ca$ as the $Q^2$-rescaling model in the range $0.2 < x < 0.35$
although there is a slight difference in the nuclear shadowing region ($x< 0.1$).
The results of the nuclear Drell-Yan ratios for other nuclei such as
C, Fe, W  are not presented since the  results are 
qualitatively similar. The results of these nuclear Drell-Yan ratios
do not show a  distinct nuclear dependence, which is consistent with 
the experimental fact.

In summary, we have put forward the APE model
which attempts to explain the nuclear effect on the parton distributions 
by the change of the transverse momentum of partons in the nuclear
environment instead of the $Q^2$-rescaling model. 
By means of the APE model with the parton recombination effect, we 
investigated the nuclear effect including the nuclear structure function
$F_2^A(x,Q^2)$ in the whole $x$ region, nuclear Drell-Yan process 
and nuclear gluon distribution. 
 The results of the nuclear structure 
function , in the small $x$ region ($x< 0.06$),
are of a little $Q^2$-dependence  but  are very sensitive to the input 
sea-quark and gluon distributions. By using the proper input parton 
distributions, we obtained reasonably good 
agreements with the experimental data of the EMC effect, nuclear Drell-Yan ratio
and nuclear gluon distributions.

In addition, the comparison  between the APE model and 
the $Q^2$-rescaling model shows,
on one hand, that the results are   
qualitatively similar since they have the same physics
origin; and on the other hand, with better description of the nuclear 
effect on sea quarks and gluons, the value of $R^{A/D}$ 
in the APE model is less 
sensitive to the choice of the free parton distributions than 
that of  the $Q^2$-rescaling
model. Our present investigation should be 
considered as  the first step for seeking for  more subtle mechanism
with the same physics origin as the $Q^2$-rescaling model but with less
uncertainties. Further, we are looking forward to
a unified description of the nuclear effect and the nuclear
shadowing within the framework of the APE model. 

\mbox{}\hspace{3cm}{ ACKNOWLEDGMENTS} 

This work was supported by National Natural Science Foundation of China and 
Natural Science Foundation of Jiangsu Province of China. One of the 
authors (J.J.Y) would like to thank Dr. S. Kumano for his correspondence
about the A-P equation subroutine, and Dr. Vogt for sending 
their computer programs for calculating the parton distributions in 
Ref. \cite {GRV}. He also expresses thanks to Prof. H. Q. Shen and Dr. S. A.  
Aruna for helpful discussions.

\newpage

\section*{Figure Captions}

\vspace{0.3cm}
\noindent
\hspace{0.65cm}Fig.1. Deep inelastic scattering in the free parton model.

\vspace{0.3cm}
\noindent
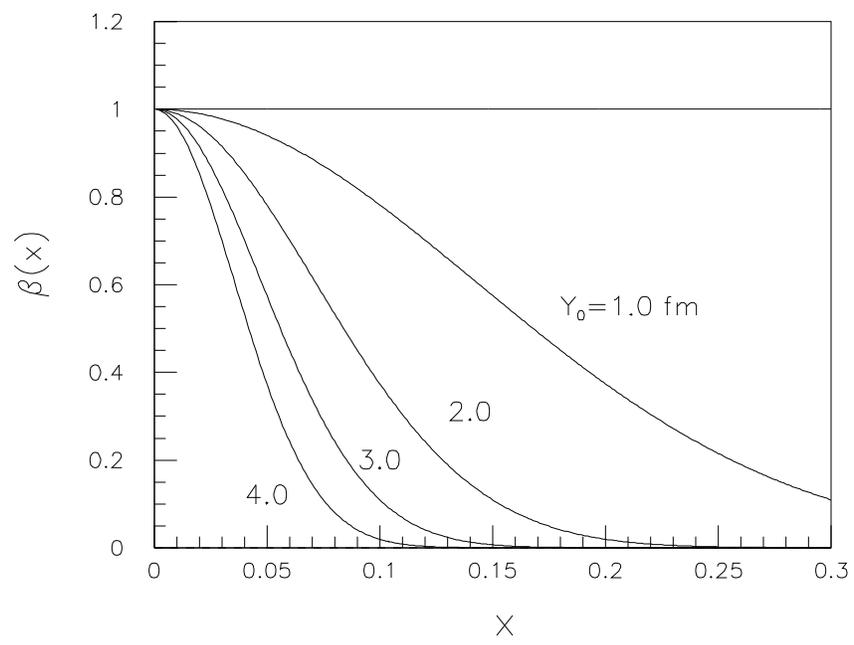
Fig.2. Momentum cutoff for leak-out partons, 
$\beta(x)=exp(-\frac {m_N^2y_0^2x^2}2)$.

\vspace{0.3cm}
\noindent
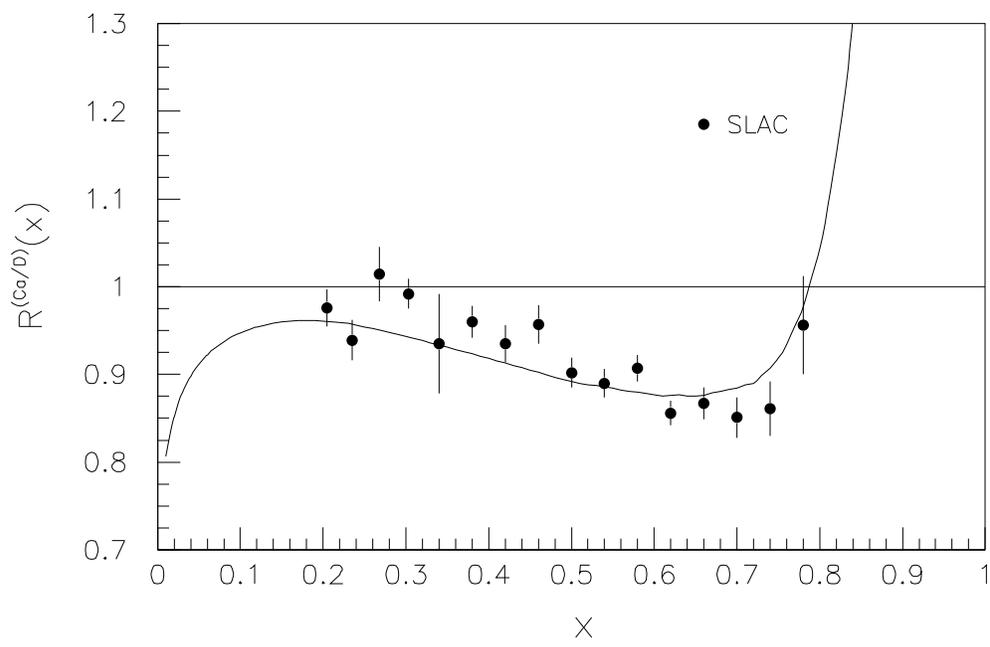
Fig.3. Comparisons with SLAC data for $^{40}Ca$ 
with Ferim motion correction. Input parton distributions are those of 
KMRS-B0 \cite {KMRS}. $Q^2=4.0GeV^2$.

\vspace{0.3cm}
\noindent
Fig.4(a-b). Comparisons of the ratio $R^{Ca/D}$ with 
 EMC-90 \cite {EMC-2} and 
NMC \cite {NMC-EMC} data by using (a) the GRV \cite {GRV} input 
and (b) the KMRS-B0 \cite {KMRS} input. $Q^2=4.0GeV^2$ and $8.0GeV^2$
for the solid and dashed curves, respectively.
   
\vspace{0.3cm}
\noindent
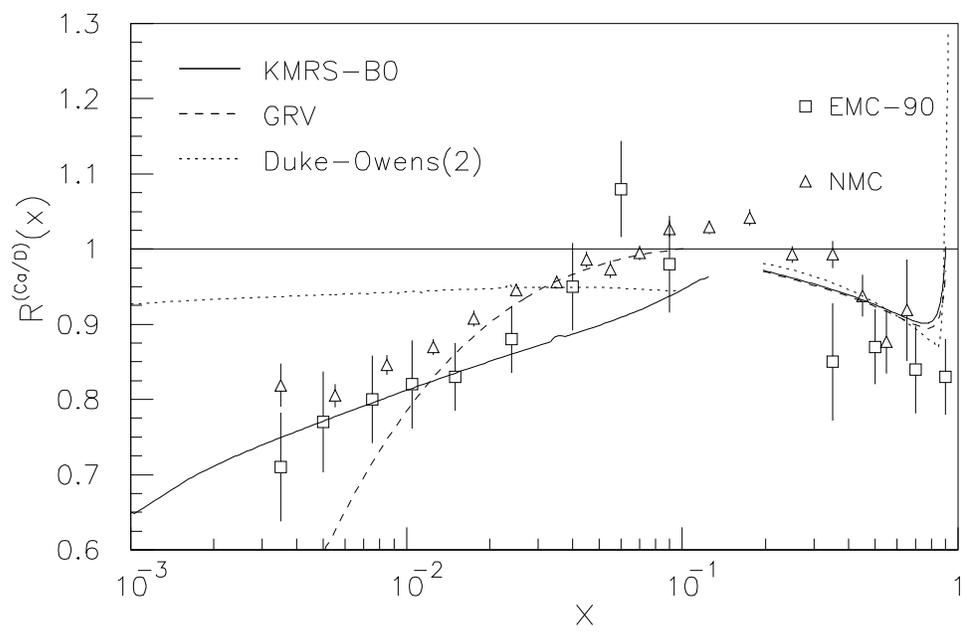
Fig.5. Comparisons of the ratio $R^{Ca/D}$ with 
 EMC-90 \cite {EMC-2} and 
NMC \cite {NMC-EMC} data by using different  input distributions.
The solid, dashed and dotted curves are for the KMRS-B0 \cite {KMRS}
,the GRV \cite {GRV} and Duke-Owens(2) \cite {D-O}, respectively.
 $Q^2=4.0GeV^2$.

\vspace{0.3cm}
\noindent
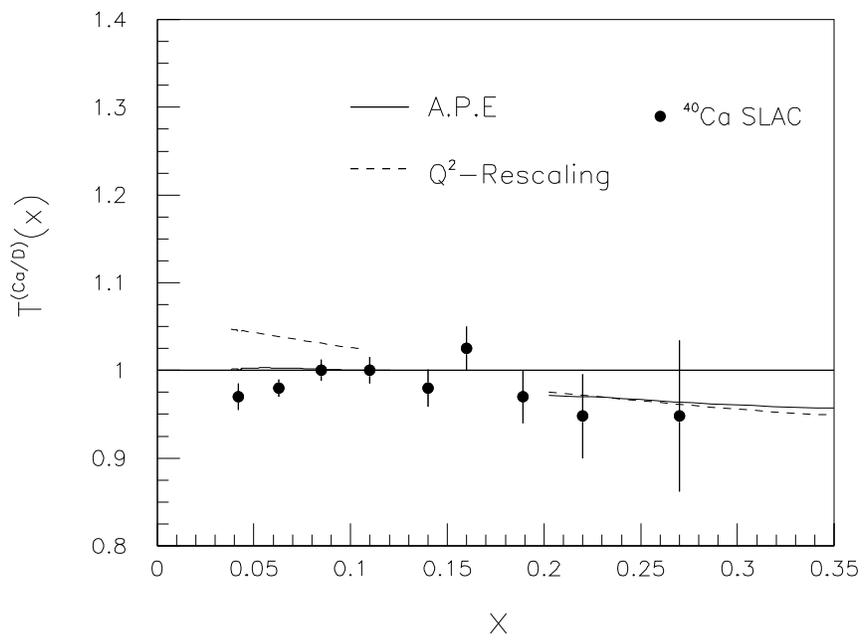
Fig.6. The nuclear Drell-Yan ratios $T^{Ca/D}(x)$ predicted by the present
model(solid  curve) and the $Q^2$-rescaling model(dashed curve) 
 are plotted versus x by using the GRV 
input distributions at $Q^2$=20 $GeV^2$.
 The experimental data are taken from the E772 Collaboration \cite {Alde}.

\vspace{0.3cm}
\noindent
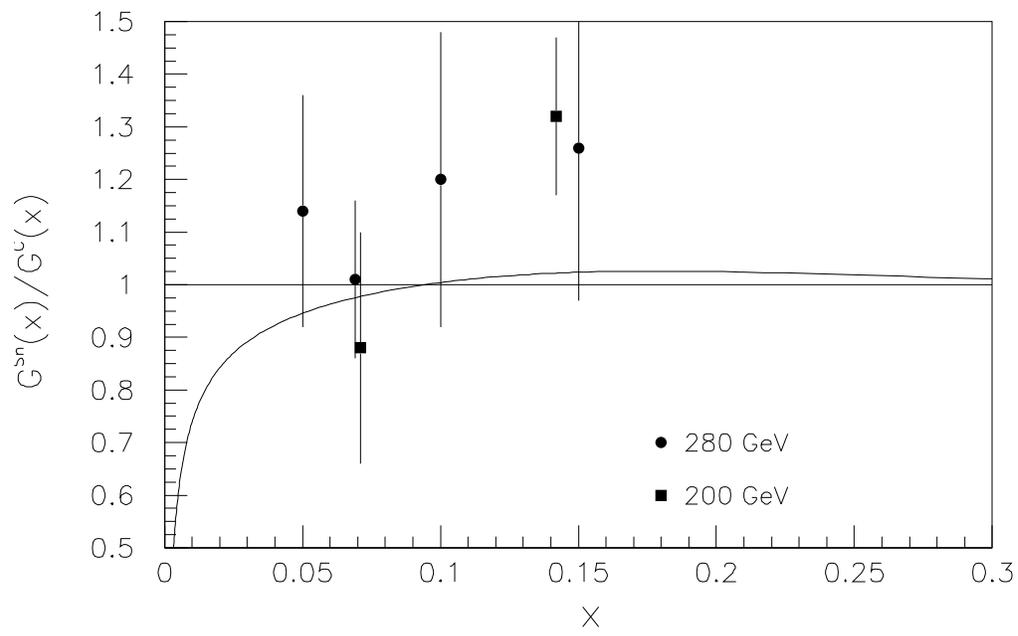
Fig.7. Comparisons of the result of $G_{Sn}(x)/G_{C}(x)$ with 
NMC data \cite {NMC} by using the KMRS-B0 
input distributions at $Q^2$=4 $GeV^2$.

\vspace{0.3cm}
\vspace{0.3cm}
\noindent
Fig.8(a-b). The ratio $R^{Ca/D}(x)$ given by the APE model 
is plotted. (a) in the linear x scale; (b) in the logarithmic  x scale.
Results by using the GRV inputs(solid curve) and the Duke-Owens(2) inputs(dashed
curve) at $Q^2$=4 $GeV^2$.
\noindent

\vspace{0.3cm}
\noindent
Fig.9(a-b). The ratio $R^{Ca/D}(x)$ given by the $Q^2$-rescaling model 
is plotted. (a) in the linear x scale; (b) in the logarithmic  x scale.
Results by using the GRV inputs(solid curve) and the Duke-Owens(2) inputs(dashed
curve) at $Q^2$=4 $GeV^2$.

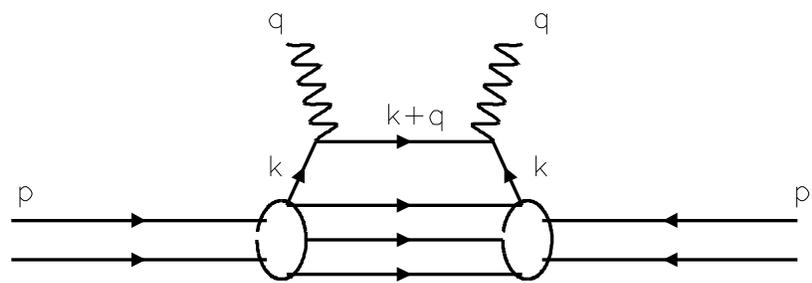
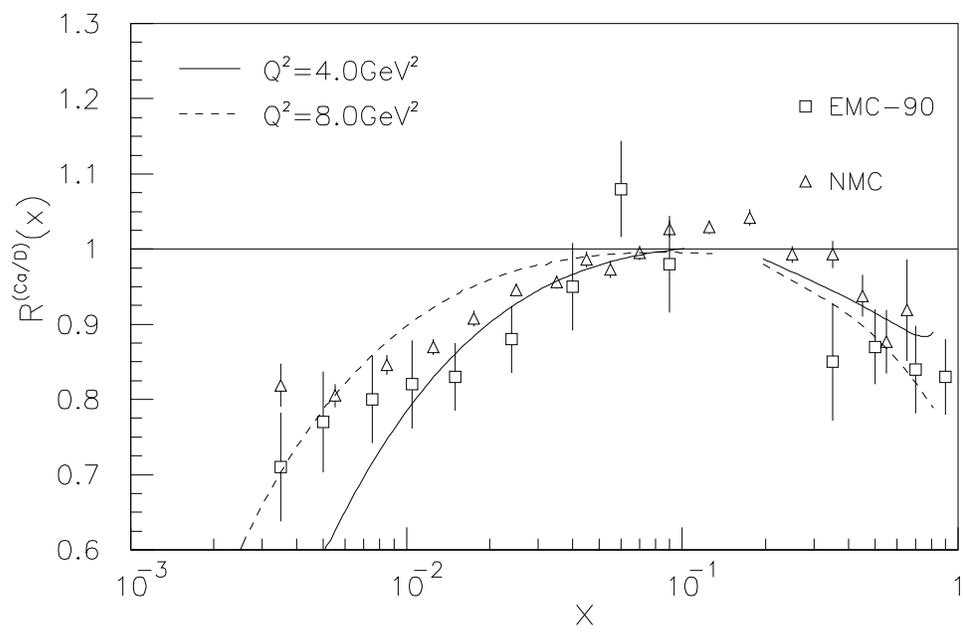
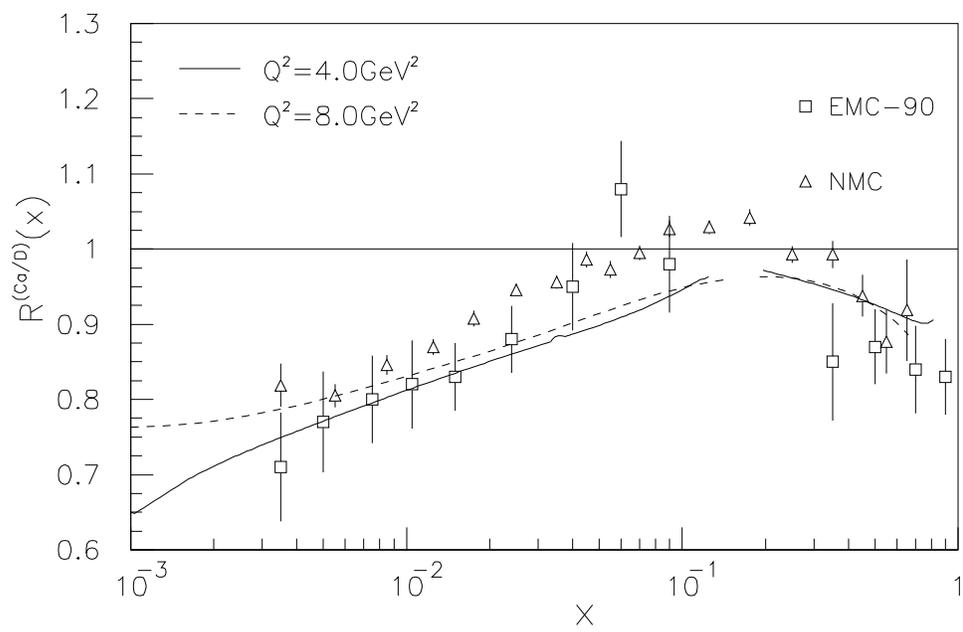
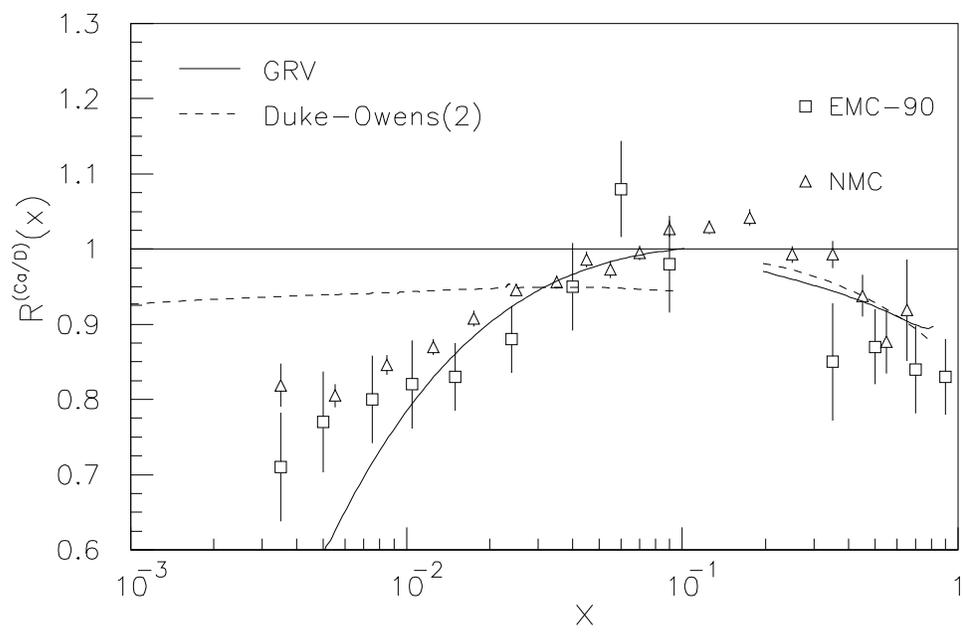
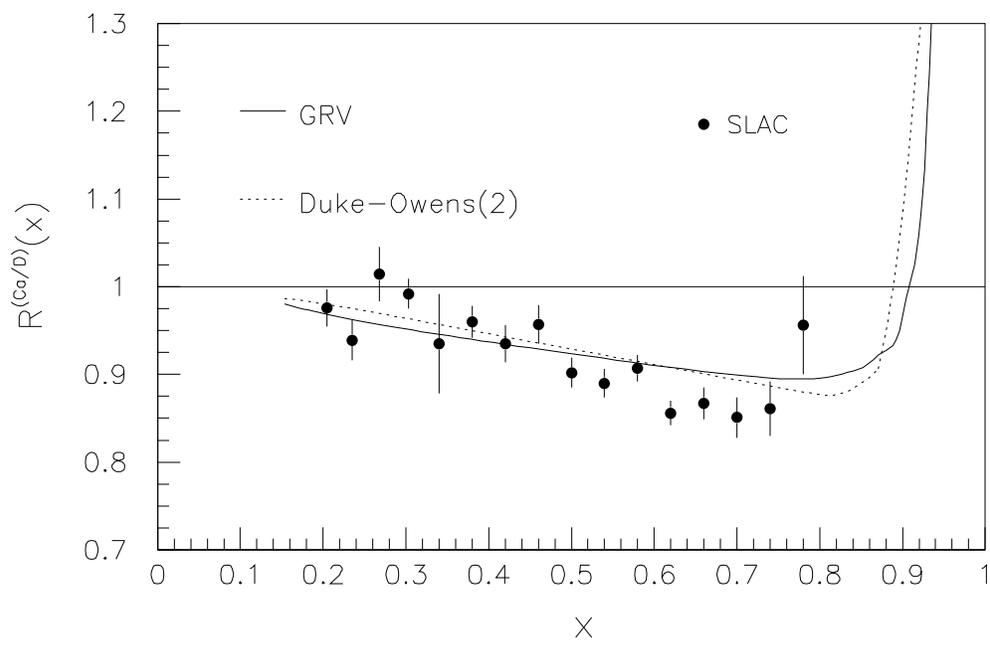
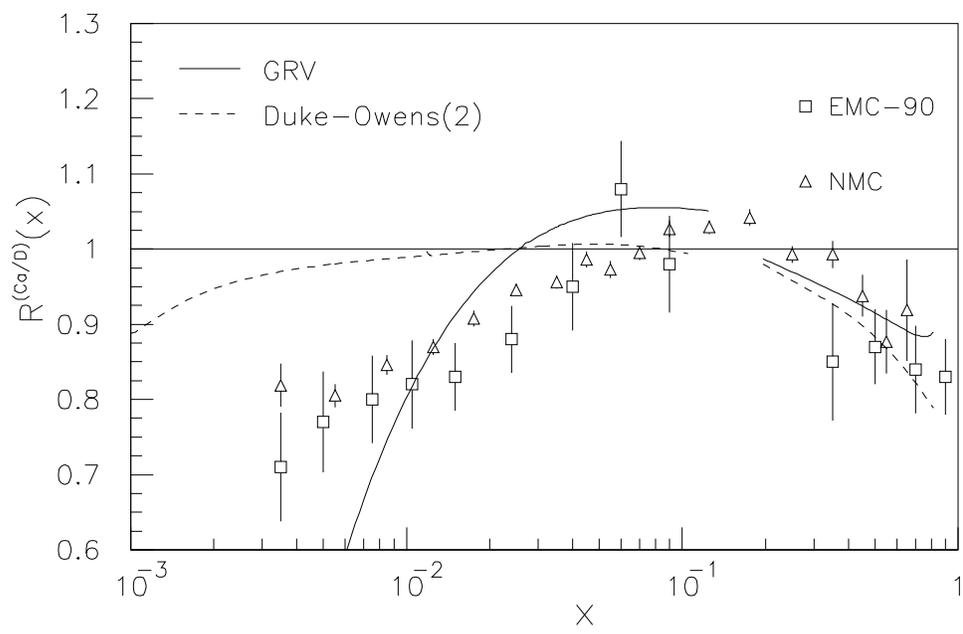
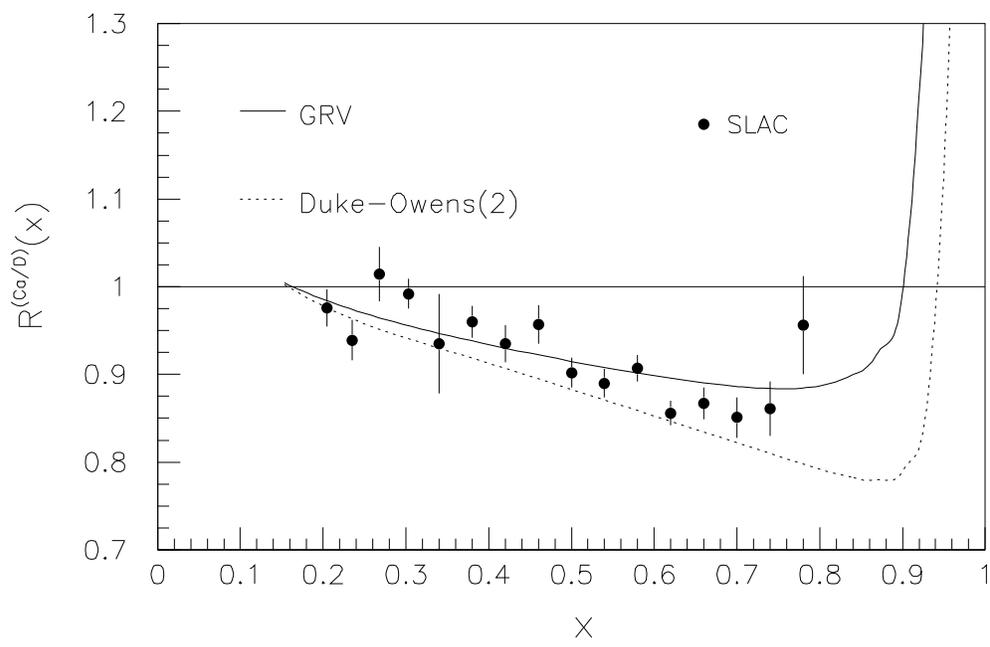
\end{document}